\documentstyle[11pt]{article}

\begin{document}

\title{ 
QCD Sum-Rule Invisibilty  of the  $\sigma$ Meson
}
\author{
T.G. Steele, Fang Shi        \\
Department of Physics,
 University of Saskatchewan, \\
Saskatoon, SK, S7N 5E2\\
Canada \\[5pt]
V. Elias        \\
 Department of Applied Mathematics,\\
 University of Western Ontario, \\
London, ON, N6A 3K7\\
Canada.
}
\maketitle
\baselineskip=14.5pt
\begin{abstract}
QCD Laplace sum-rules for light-quark $I=0,1$ scalar currents are used to investigate
candidates for  the lightest $q\bar q$ scalar mesons.  The theoretical 
predictions for the sum-rules include  instanton contributions which split the degeneracy between the
$I=0$ and  $I=1$ channels.  The self-consistency of the theoretical predictions is verified through
a H\"older inequality analysis, confirming the existence of an effective instanton contribution to the
continuum.  The sum-rule analysis indicates that the $f_0(980)$ and $a_0(1450)$ should be interpreted
as the lightest $q\bar q$ scalar mesons.  This apparent  decoupling of the $f_0(400-1200)$ (or $\sigma$)
and $a_0(980)$ from the quark scalar currents suggests a non-$q\bar q$ interpretation of these resonances.
\end{abstract}
\baselineskip=17pt
\section{Field-Theoretical Content of the Sum-Rule}
The nature  of the scalar mesons is a challenging problem in hadronic physics.  In particular,
a variety of interpretations exist for the lowest-lying isoscalar resonances [$f_0(400-1200)$, $f_0(980)$,
$f_0(1370)$, $f_0(1500)$] and  isovector resonances [$a_0(980)$, $a_0(1450)$] listed by the Particle Data 
Group (PDG) \cite{PDG}.   In particular,  interpreting  the
$f_0(400-1200)$ (or $\sigma$) is particularly significant because of its possible interpretation as the $\sigma$ meson of chiral 
symmetry breaking.
In this paper we will summarize and extend previous work \cite{first_paper}  which used 
QCD Laplace sum-rules to study the various possibilities for the 
lowest-lying, non-strange quark scalar mesons.  

QCD sum-rules probe hadronic properties  through correlation functions  of appropriately chosen
currents.   In  
 the $SU(2)$ flavour limit $m_u=m_d\equiv m$, the non-strange-quark $I=0,1$ scalar mesons are
studied via the scalar-current correlation function:
\begin{eqnarray}
& &J_I(x)=\frac{m}{2}\left[ \bar u(x)u(x)+ \left(-1\right)^I \bar d(x)d(x)\right ]\quad , \quad I=0,1
\label{scalar_current}\\
& &
\Pi_I\left(Q^2\right)=i\int d^4x\,e^{iq\cdot x}\langle O\vert T J_I(x)J_I(0)\vert O \rangle
\label{correlation_function}
\end{eqnarray} 
Laplace sum-rules,
which  exponentially suppress  
the high-energy region, are obtained  by applying the Borel transform operator $\hat B$ 
to  the appropriately-subtracted dispersion relation satisfied by (\ref{correlation_function}) \cite{SVZ}:
\begin{equation}
{\cal R}^I_0(\tau)\equiv \frac{1}{\tau}\hat B\left[\Pi_I\left(Q^2\right)\right]
=\frac{1}{\pi}\int\limits_0^\infty Im\Pi_I(t) e^{-t\tau}\,dt
\label{basicsr}
\end{equation}

To leading order in the quark mass, the theoretical prediction for ${\cal R}_0^I$ incorporates two-loop
$\overline{{\rm MS}}$ scheme perturbative corrections \cite{4-loop}, infinite correlation-length non-perturbative 
vacuum effects parametrized by the QCD condensates \cite{SVZ,condensates}, and finite-correlation length non-perturbative effects
of instantons in the instanton liquid model \cite{instanton_sr,ins_liquid}:
\begin{eqnarray}
{\cal R}^I_0(\tau)&=&\frac{3m^2}{16\pi^2\tau^2}\left(   
1+4.821098 \frac{\alpha}{\pi}
\right)
+m^2\left(
\frac{3}{2}\langle m\bar q q\rangle 
+\frac{1}{16\pi}\langle \alpha G^2\rangle
+\pi\langle{\cal O}_6\rangle \tau
\right)
\nonumber
\\
& & \qquad +\left(-1\right)^Im^2
{3\rho_c^2\over{16 \pi^2\tau^3}} e^{-\frac{\rho_c^2}{2\tau} }
\left[   
  K_0\left( \frac{\rho_c^2}{2\tau} \right) +
       K_1\left(\frac{\rho_c^2}{2\tau}  \right)
\right]
\label{sr}
\end{eqnarray}
where the quantity $\rho=1/(600\,{\rm MeV})$ is the mean instanton size in the instanton liquid model \cite{ins_liquid}.
The only theoretical source of  isospin-breaking effects in (\ref{sr}) are instantons, which
are known to have non-trivial contributions for only the scalar and pseudoscalar correlation functions.
 
We have used $SU(2)$ symmetry for the dimension-four quark condensate 
contributions to (\ref{sr})
({\it i.e.} 
$\langle \bar u u\rangle= \langle \bar d d\rangle\equiv  \langle \bar q q\rangle$).
The quantity  $\langle {\cal O}_6\rangle$ denotes the dimension-six quark condensates 
for which the  vacuum saturation hypothesis \cite{SVZ} provides  a reference value
\begin{equation}
\langle{\cal O}_6\rangle=-f_{vs}\frac{88}{27}\alpha
\langle \bar q q\bar q q\rangle
=-f_{vs}5.9\times 10^{-4} {\rm GeV}^6 
\label{o61}
\end{equation}
where $f_{vs}=1$ for exact vacuum saturation.  Larger values 
of effective dimension-six
operators found in \cite{dimsix} imply that $f_{vs}$  could be as 
large as $2$, suggesting  a central value $f_{vs}=1.5$.
The quark condensate is determined by the GMOR (PCAC) relation, 
and  the gluon condensate is given by
\cite{dimsix}
\begin{equation}
\langle \alpha G^2\rangle=\left(0.045\pm 0.014\right)\,{\rm GeV^4}
\label{aGG}
\end{equation}

Renormalization group improvement of (\ref{sr}) implies that $\alpha$ and $m$ are
 running quantities evaluated at the mass scale $Q=\frac{1}{\sqrt{\tau}}$  in 
the $\overline{{\rm MS}}$ scheme.  
We use $\Lambda_{\overline{MS}}\approx 300\,{\rm MeV}$ for three active flavours,
consistent with current estimates of $\alpha(M_\tau)$ and matching conditions through the charm threshold \cite{PDG,run_alpha}.

Phenomenological analysis of the sum-rule (\ref{basicsr}) proceeds through the resonance plus continuum model
\cite{SVZ}
\begin{equation}
Im\Pi_I(t)=Im\Pi_I^{res} +\theta\left(t-s_0\right)Im\Pi_I^{QCD}(t)
\label{basic_phenom}
\end{equation}
where $Im\Pi_I^{res}$ denotes the resonance contributions, and $Im\Pi_I^{QCD}$ represents the theoretically-determined 
QCD continuum occurring above the continuum threshold $s_0$.  Defining these continuum contributions as
\begin{equation}
c_0^I\left(\tau,s_0\right)=
\frac{1}{\pi}\int\limits_{s_0}^\infty Im\Pi_I^{QCD}(t) e^{-t\tau}\,dt
\end{equation}
leads to a revised sum-rule which isolates the theoretical and phenomenological (resonance) contributions:
\begin{equation}
{\cal S}^I_0\left(\tau,s_0\right)\equiv {\cal R}^I_0(\tau)-c_0^I\left(\tau,s_0\right)
=\frac{1}{\pi}\int\limits_0^{s_0} Im\Pi_I^{res}(t) e^{-t\tau}\,dt
\label{res_sr}
\end{equation}

Traditionally, only the perturbative contributions are included in the continuum.
However,  the $Q^2$ analytic structure of the instanton contributions to $\Pi_I^{inst}(Q^2)$ implies the existence
of an imaginary part ${\rm Im} \Pi_I^{inst}(t)$
which leads to the following instanton continuum contribution \cite{instanton_continuum}:
\begin{equation}
c^I_{0_{inst}}\left(\tau,s0\right)=\frac{1}{\pi}\int\limits_{s_0}^\infty Im\Pi_I^{inst}(t) e^{-t\tau}\,dt
=
\left(-1\right)^{I+1}\frac{3m^2}{8\pi}
\int\limits_{s_0}^\infty
t J_1\left(\rho_c\sqrt{t}\right)Y_1\left(\rho_c\sqrt{t}\right)\,dt
\label{ins_continuum}
\end{equation}
where $J_n(x)$ and $Y_n(x)$ denote Bessel functions.
The instanton continuum contribution has been ignored in previous applications of instanton effects in sum-rules.  
It should be noted that  this formulation of the instanton effects leads to improved IR behaviour when integrating 
over the instanton density  because  (\ref{ins_continuum}) approaches zero in the limit $\rho\rightarrow 0$.

\section{H\"older Inequality Constraints} 
In the phenomenological analysis of QCD sum-rules, the behaviour of ${\cal S}_0(\tau,s_0)$ as a function of 
Borel-parameter $\tau$ is used
to extract the phenomenological resonance parameters through (\ref{res_sr}),  raising the difficult question of the $\tau$ 
region where the theoretical prediction  ${\cal S}_0(\tau,s_0)$ is valid \cite{SVZ}.
This question can be addressed via H\"older inequalities, which must be upheld if Laplace
sum-rules are to be consistent with the physically-required positivity of ${\rm Im}\Pi_I^{res}(t)$
within the integrand of  (\ref{res_sr}) \cite{holder}:  
\begin{equation}
\frac{{\cal S}^I_0[\omega\tau+(1-
\omega)\delta\tau,s_0]}{\left({\cal S}^I_0[\tau,s_0]\right)^\omega
\left({\cal S}_0^I[\tau+\delta\tau,s_0]\right)^{1-\omega}} \le 1
\quad ,\quad  \forall ~0\le \omega \le 1
\label{rat2a}
\end{equation}
Provided that $\delta\tau$ is reasonably small ($\delta\tau\approx 0.1\,GeV^{-2}$
appears to be sufficient \cite{holder}),
these inequalities are  insensitive to the choice of $\delta\tau$,
permitting a simple analysis of the inequality as a function of the Borel-parameter $\tau$.

The scalar-channel sum-rules satisfy the inequality in a fashion qualitatively similar to other channels \cite{holder}, 
supporting  the 
self-consistency of the theoretical predictions.  The instanton continuum (\ref{ins_continuum})   
is crucial to this agreement.
Regions of validity in which the sum-rules satisfy the inequality (\ref{rat2a}) are
\begin{eqnarray}
0.3\,{\rm GeV}^{-2}\le \tau\le 1.7\,{\rm GeV}^{-2}~,~s_0>3\,{\rm GeV}^2\quad (I=0)
\label{f0_ineq}
\\
0.3\,{\rm GeV}^{-2}\le \tau\le 1.1\,{\rm GeV}^{-2}~,~s_0>3\,{\rm GeV}^2\quad (I=1)
\label{a0_ineq}
\end{eqnarray}

\section{Phenomenological Analysis}
The sum-rule predictions of the properties of the lowest-lying $I=0,1$ quark scalar resonances can now
be studied through  (\ref{res_sr}).  Since the resonances could have a substantial width, it is necessary to extend the narrow width approximation 
traditionally used in sum-rules.   A flexible and numerically simple technique  is 
to build up the resonance shape using $n$
 unit-area square 
pulses  \cite{first_paper,width}
\begin{eqnarray}
\frac{1}{\pi}{\rm Im}\Pi^{(n)}(t) &=&  \frac{2}{n\pi} \sum_{j=1}^n \sqrt{ \frac{n-j+f}{j-f}} P_M \left[
t, \sqrt{\frac{n-j+f}{j-f}} \; \Gamma \right]
\label{n-pulse}
\\
P_M(t,\Gamma)&=&\frac{1}{2M\Gamma}\left[\Theta(t-M^2+M\Gamma)-\Theta(t-M^2-M\Gamma)\right]
\label{square_pulse}
\end{eqnarray}
A single square pulse models a broad nearly structureless contribution (such as a broad light $\sigma$) to ${\rm Im}\Pi(t)$,
while a Breit-Wigner resonance of a particle of mass $M$ and width $\Gamma$  can be expressed as a sum of several square pulses.
The quantity $f$ can be fixed by normalizing the area of the n-pulse approximation to unity.

We begin the phenomenological analysis with the $4$-pulse approximation (\ref{n-pulse})
to ${\rm Im}\Pi^{res}(t)$ so that (\ref{basicsr}) becomes
\begin{eqnarray}
& &\frac{1}{\pi}{\rm Im}\Pi_I^{res}=F^2M^4\frac{1}{\pi}{\rm Im}\Pi^{(4)}(t)\quad ,\quad
{\cal S}^I(\tau,s_0)=F^2M^4e^{-M^2\tau}W_4(M,\Gamma,\tau)\label{basic_fit}\\
& &W_4(M,\Gamma,\tau)=
\frac{2}{4\pi} \sum_{j=1}^4 \frac{1}{M\Gamma\tau}\sinh\left[M\sqrt{\frac{4-j+f}{j-f}}\,\Gamma\tau\right]
\label{n-pulse-W}
\end{eqnarray}
where $F$ is the strength with which the scalar current couples the vacuum to the resonance. 
The free parameters in this expression, the resonance-related quantities $F$, $M$, $\Gamma$ and the continuum-threshold 
$s_0$, can be extracted from a fit to the $\tau$ dependence of the theoretical expression
${\cal S}^I(\tau,s_0)$.  This is done by minimizing the $\chi^2$ defined by
\begin{equation}
\chi^2=\frac{1}{N}\sum_{j=1}^N\frac{\left[{\cal S}^I\left(\tau_j,s_0\right)-F^2M^4e^{-M^2\tau_j}W_4(M,\Gamma,\tau_j)\right]^2}{\epsilon(\tau_j)^2}
\label{chi2}
\end{equation}  
where the sum is over evenly spaced, discrete $\tau$ points in the ranges (\ref{f0_ineq},\ref{a0_ineq}) consistent with the
H\"older inequality.  The weighting factor $\epsilon$ used for the evaluation of (\ref{chi2}) is
$\epsilon(\tau)=0.2{\cal S}^I(\tau,s_0)$.
This 20\% uncertainty has the desired property of being dominated by the continuum at low $\tau$ and power-law corrections at large
$\tau$.  Other choices of the $0.2$ prefactor in $\epsilon$ would simply rescale the $\chi^2$, so its choice has no effect on
the values of the $\chi^2$-minimizing parameters.

In the $\chi^2$ minimization, the quark mass parameter $\hat m$ is now absorbed into the quantity $a=F^2M^4/\hat m^2$. 
The best-fit parameters are subjected to a Monte-Carlo simulation which includes  the 
parameter ranges $1\le f_{vs}\le 2$, a 15\% variation in the instanton size $\rho$, and a simulation of continuum and OPE truncation
uncertainties.  This  results in the 90\% confidence level results for the best-fit parameters shown in   Table \ref{mc_tab}.  Decreasing the number of pulses (to simulate a structureless resonance) does not alter the $\chi^2$, and only leads to a rescaling of $\Gamma$.  Two-resonance models recover the
single-resonance results in Table \ref{mc_tab}, so there is no evidence of a hidden light resonance in either of the channels.

\begin{table}[hbt]
\begin{tabular}{||c|c|c|c|c||}\hline\hline
$I$ & $M~(GeV)$ & $s_0~(GeV^2)$ & $a~(GeV^4)$ & $\Gamma ~(GeV)$
\\ \hline\hline
0  &  $1.00\pm 0.09$ & $3.7\pm 0.4$ & $0.08\pm 0.02$ & $0.19\pm 0.14$ 
\\ \hline
1 & $1.55\pm 0.11$  & $5.0\pm 0.7$  &$0.17\pm 0.04$ & $0.22\pm 0.11$
\\\hline\hline
\end{tabular}
\caption{{ \it Results of the Monte-Carlo simulation of 90\% confidence-level uncertainties 
 for the resonance parameters and continuum threshold for the $I=0,1$ channels.}}
\label{mc_tab}
\end{table}

Thus we conclude that a QCD sum-rule analysis is consistent with the interpretation of the $f_0(980)$ and $a_0(1450)$ as the lightest
non-strange quark scalar mesons.  A light $\sigma$ meson [$f_0(400-1200)$] and the $a_0(980)$ appear to be decoupled from the
quark scalar currents, suggesting a non-$q\bar q$ interpretation of these resonances.

The authors gratefully acknowledge research funding from the Natural Sciences and Engineering Research Council of Canada (NSERC).


\begin{thebibliography}{99}
\bibitem{PDG}  Particle data Group, C. Caso  {\it et al} Eur. Phys. J.  {\bf C3}, 1 (1998).

\bibitem{first_paper}  V. Elias, A. H. Fariborz, Fang~Shi, T.G.~Steele, Nucl.Phys. {\bf A633}, 279 (1998).

\bibitem{SVZ}
M. A. Shifman, A. I. Vainshtein, and V. I. Zakharov: Nucl.
Phys. {\bf B147}, 385 (1979).


\bibitem{4-loop} K.G. Chetyrkin, Phys. Lett. {\bf B390}, 309 (1997);
S.G. Gorishny, A.L.~Kataev, S.A.~Larin,L.R.~Surguladze, Phys. Rev. {\bf D43}, 1633 (1991).


\bibitem{condensates} 
E. Bagan, J.I. LaTorre, P. Pascual, Z. Phys. {\bf C32}, 43 (1986).

\bibitem{instanton_sr}
A. E. Dorokhov, S. V.Esaibegian, N. I. Kochelev, N. G. Stefanis, J. Phys. {\bf G23}, 643 (1997).

\bibitem{ins_liquid} E. V. Shuryak, Nucl. Phys. {\bf B214}, 237 (1983). 


\bibitem{dimsix} C.A. Dominguez, J. Sola, Z. Phys. {\bf C40}, 63 (1988);
V. Gimenez, J. Bordes, J.A. Penarrocha, Nucl. Phys. {\bf B357}, 3 (1991).

\bibitem{run_alpha} K.G. Chetyrkin, B.A. Kniehl, M. Steinhauser, Phys. Rev. Lett. {\bf 79}, 2184 (1997); 
T.G. Steele, V. Elias, Mod. Phys. Lett. {\bf A13}, 3151 (1998). 

\bibitem{instanton_continuum} V. Elias, Fang Shi, T.G. Steele, J. Phys. {\bf G24}, 267 (1998); A.S. Deakin, V. Elias, 
Ying Xue, N.H. Fuchs, Fang Shi, T.G. Steele, Phys. Lett. {\bf B418}, 223 (1998). 


\bibitem{holder} M. Benmerrouche, G. Orlandini, T.G. Steele, Phys. Lett. {\bf B356}, 573 (1995).

\bibitem{width} V. Elias, A.H.~Fariborz, M.A.~Samuel, Fang~Shi, T.G.~Steele,
Phys. Lett. {\bf B412}, 131 (1997).

\end{thebibliography}
\end{document}